\documentclass[prd,onecolumn, nofootinbib, 11pt]{revtex4}  
\usepackage{graphicx}
\usepackage{amsmath}
\usepackage{amsfonts}
\usepackage{amssymb}
\usepackage{bm}
\usepackage{appendix}
\usepackage{mathtools}
\usepackage{comment}
\usepackage{bbold}
\usepackage{color}
\usepackage{slashed}
\usepackage{subfigure}
\usepackage{setspace}
\usepackage{enumitem}
\usepackage{longtable}
\usepackage{wasysym}
\usepackage[usenames,dvipsnames]{xcolor}
\usepackage{bm}
\usepackage{multirow}
\usepackage{changepage}
\usepackage[letterpaper, margin=.8in, top=0.85in, bottom=0.85in]{geometry}

\pdfoutput=1

\newcommand{\iD}{iD_\perp}
\newcommand{\D}{D_\perp}
\newcommand{\p}{\partial_\perp}

\begin{document}

\title{Hilbert Series and Operator Basis for NRQED and  NRQCD/HQET}
\author{Andrew Kobach}
\affiliation{Physics Department, University of California, San Diego, La Jolla, CA 92093, USA}
\author{Sridip Pal}
\affiliation{Physics Department, University of California, San Diego, La Jolla, CA 92093, USA}

\date{\today}

\begin{abstract}
We use a Hilbert series to construct an operator basis in the $1/m$ expansion of a theory with a nonrelativistic heavy fermion in an electromagnetic (NRQED) or color gauge field (NRQCD/HQET).  We present a list of effective operators with mass dimension $d\leq 8$.  Comparing to the current literature, our results for NRQED agree for $d\leq 8$, but there are some discrepancies in NRQCD/HQET at $d=7$ and 8. 
\end{abstract}

\maketitle

\linespread{1}

\section{Introduction}
 An operator basis for an effective field theory is a set containing all operators that give rise to different scattering matrix elements, invariant under relevant symmetries of the theory. The Hilbert series\footnote{The Hilbert series is a generic concept, defined on any graded vector space. In our context, it is defined over the ring of operators of the effective theory under consideration, as in Refs.~\cite{Henning:2015daa, Henning:2015alf}. Our working definition is given in Section~\ref{secHS}.} can be used as a tool for enumerating the elements of an operator basis for effective field theories. 
With it, one can impose symmetry requirements~\cite{Feng:2007ur, Jenkins:2009dy, Hanany:2010vu} and account for redundancies between operators coming from the equations of motion and integration by parts~\cite{Henning:2015daa, Henning:2015alf}.    
So far, the focus has been on relativistic theories, e.g., especially the effective theory of the Standard Model~\cite{Henning:2015alf, Lehman:2015via, Lehman:2015coa}, and nonrelativistic effective theories have been unexplored using Hilbert-series methods.

We examine the specific effective theories of a single nonrelativistic fermion in an electromagnetic field or color field.  These effective theories are called non-relativistic QED (NRQED) and non-relativistic QCD (NRQCD), respectively.  
NRQCD is the same theory described by the heavy quark expansion (HQET)~\cite{Manohar:1997qy}, and we refer to this theory as NRQCD/HQET.  
One can construct a list of effective operators, where operators are suppressed by the appropriate powers of $1/m$ (where $m$ is the fermion's mass), in the rest frame of the heavy fermion without loss of generality. 
These two effective theories have been used extensively over the past few decades.  For example, NRQED was originally formulated in Ref.~\cite{Caswell:1985ui}, where higher-dimensional operators were listed by the authors of Refs.~\cite{Kinoshita:1995mt, Manohar:1997qy, Hill:2012rh}, and is used to explore the proton radius puzzle (see, for example, Ref.~\cite{Hill:2017wzi}, and references therein). 
NRQCD/HQET is a tool that can be used to extract the value of $|V_{cb}|$ in inclusive semileptonic $B$ decays, and the higher-order terms in $1/m$ have been discussed in Refs.~\cite{Mannel:1994kv, Manohar:1997qy, Mannel:2010wj, Gunawardana:2017zix}.  These high-order terms in $1/m$ may become important when analyzing the high-luminosity data from the upcoming Belle-II experiment~\cite{Gambino:2016jkc, Heinonen:2014dxa}.

There is currently no disagreement in the literature regarding the number of NRQED operators up to and including order $1/m^4$, and the Hilbert series we construct for NRQED agrees with the results in Refs.~\cite{Caswell:1985ui, Kinoshita:1995mt, Manohar:1997qy, Hill:2012rh}.  
Also, our results for NRQCD/HQET agree with those in Refs.~\cite{Mannel:1994kv, Manohar:1997qy} up to order $1/m^3$, but we find discrepancies with other analyses at $1/m^3$ and $1/m^4$.  
Specifically, we count 11 operators at $1/m^3$ (as does Ref.~\cite{Manohar:1997qy}), and 25 operators at $1/m^4$.  
However, at order $1/m^3$, Ref.~\cite{Dassinger:2006md} says there are 5, and Refs.~\cite{Mannel:2010wj, Gunawardana:2017zix} claim there are 9.  At order $1/m^4$, Refs.~\cite{Mannel:2010wj, Gunawardana:2017zix} claim there are 18 operators.  
The differences between our results and those found in Refs.~\cite{Mannel:2010wj, Gunawardana:2017zix} could be explained by there being two symmetric $SU(3)$ color singlets for operators with two gauge bosons.  
We discuss this further in Section~\ref{secNRQCD}.

\section{Effective Theory for a Nonrelativistic Fermion}

We consider a system where the relevant dynamics  of a massive fermion in an external, dynamical, gauge field occurs at energy scales well below the rest mass, $m$, of a fermion.\footnote{Physical theories are diffeomorphic, so if an operator is zero in one reference frame, it is zero in all other frames. Therefore, we choose to work in the rest frame of the nonrelativistic fermion for the purposes of enumerating operators and constructing a basis. } The following effective Lagrangian can be used to describe such a system with a heavy fermion:
\begin{equation}
\label{lagrangian}
\mathcal{L} = \psi^\dagger iD_t \psi + \displaystyle \sum_{k=1}^\infty   c_k \psi^\dagger \mathcal{O}_k \psi .
\end{equation}
Here, $\psi$ is a two-component Pauli spinor, $c_k$ is a coupling constant, and $\mathcal{O}_k$ are Hermitian operators, suppressed by the appropriate powers of $1/m$.  
All operators $\mathcal{O}_k$ must be rotationally and translationally invariant, contain either zero or one spin vector $s^i$, and are built from time and spatial components of covariant derivatives, i.e., $iD_t$ and $iD_\perp$, respectively.

Listing all operators that satisfy only these conditions leads to over counting, since some operators can be related to others via integration by parts or relations associated with the equations of motion.  
In particular, operators with derivatives that act on $\psi^\dagger $ can be related to other operators with derivatives that act on $\psi$ by integrating by parts: 
\begin{eqnarray}
\psi^\dagger i\overleftarrow{\partial_t} \mathcal{O} \psi + \psi^\dagger i\partial_t \mathcal{O} \psi &=& 0 , \\
\psi^\dagger i\overleftarrow{\partial_\perp} \mathcal{O} \psi + \psi^\dagger i\partial_\perp \mathcal{O} \psi &=& 0 ,
\end{eqnarray}
where $\mathcal{O}$ is some operator.  
Also, the equation of motion for $\psi$ is
\begin{equation}
iD_t \psi + \displaystyle \sum_{k=1}^\infty   c_k  \mathcal{O}_k \psi = 0 .
\end{equation}
Therefore, if $D_t$ acts on $\psi$, it can be replaced by a series of operators, all at higher powers in $1/m$:
\begin{equation}
\psi^\dagger \mathcal{O} iD_t \psi = - \displaystyle \sum_{k=1}^\infty   c_k \psi^\dagger \mathcal{O} \mathcal{O}_k \psi ,
\end{equation}
where $\mathcal{O}$ is some Hermitian operator.  
A similar argument holds for the equations of motion associated with $\psi^\dagger$.  
There are also equations of motion associated with the external gauge field.  We refer to antisymmetric combinations of covariant derivatives as $E^i = \frac{i}{g}[D_t, D^i_\perp]$ and $B^i = -\frac{i}{2g} \epsilon^{ijk} [D_{\perp j}, D_{\perp k}]$.  The equations of motion for $E^i$ and $B^i$ are Maxwell's equations, or its non abelian version:
\begin{eqnarray}
\bm{D}_{\perp} \cdot \bm{E} &=& \rho ,  \\
\bm{D}_{\perp} \cdot \bm{B} &=& 0 , \\
\label{gaugeEOM}
\bm{D}_{\perp} \times \bm{E} &=& -D_{t} \bm{B} ,  \\
\bm{D}_{\perp} \times \bm{B} &=& \bm{J} + D_{t} \bm{E} ,
\end{eqnarray}
where $\rho$ and ${\bm J}$ are the external charge and current densities, respectively.  
In summary, correct enumeration of operators, when accounting for redundancies associated with integration by parts and the equations of motion,  amounts to removing:~(1) total derivatives, (2) those of the form $\psi^\dagger i\overleftarrow{D}_t \mathcal{O}\psi$ and $\psi^\dagger \mathcal{O} iD_t \psi$, (3) those with $\bm{D}_{\perp} \cdot \bm{B}$, and (4) those with either $\bm{D}_{\perp} \times \bm{E}$ or $D_{t} \bm{B}$.

More symmetry is expected in a theory with a nonrelativistic fermion, such as reparameterization invariance~\cite{Luke:1992cs, Manohar:1997qy} or residual Lorentz symmetry~\cite{Heinonen:2012km}. 
Imposing this invariance would require establishing relationships between the coefficients of operators at different orders in $1/m$.  For this work, however, we focus only on a rotationally- and translationally-invariant theory, since this can be readily encoded into a Hilbert series.  In the particular examples of NRQED and NRQCD/HQET, invariance under parity and time reversal transformations are also expected, since the underlying theories are invariant under parity and time reversal, which we discuss in Sections~\ref{secNRQED} and~\ref{secNRQCD}.

\section{Hilbert series for a nonrelativistic theory}
\label{secHS}

The Hilbert series can be used to count the number of invariants under a group transformation, utilizing the plethystic exponential, defined as
\begin{eqnarray}
\label{pebosons}
PE^\text{bosons}_\phi &\equiv& \exp\left[ \displaystyle \sum_{n=1}^\infty \frac{\phi^n}{n} \chi_{R}(z_1^n, z_2^n,...,z_k^n) \right] , \\
\label{pefermions}
PE^\text{fermions}_\psi &\equiv& \exp\left[ \displaystyle \sum_{n=1}^\infty \frac{(-1)^{n+1}\psi^n}{n} \chi_{R}(z_1^n, z_2^n,...,z_k^n) \right] .
\end{eqnarray}
Here, $\chi_{R}$ is the character of the representation $R$ of group $G$ of rank $k$, $\phi$ and $\psi$ are spurions (complex numbers taken to have modulus less than unity) corresponding to the field associated with the representation $R$, and the $z_i$'s are complex numbers with unit modulus (called fugacities) that parameterize the maximal torus of $G$.  
The plethystic exponentials are defined so as to ensure, if Taylor expanded in $\phi$ or $\psi$, that the $n$th power of $\phi$ or $\psi$ will have a coefficient equal to the character of symmetric (in the case of bosonic statistics) or antisymmetric (in the case of fermionic statistics) tensor products, constructed out of representation $R$, $n$ times. 
The Hilbert series that counts the total number of group invariants is generated by performing the following integral (often called the Molien-Weyl formula):
\begin{equation}\label{eq:hilbertdef}
HS = \oint [d\mu]_G  ~PE_x,
\end{equation}
where the contour integral is done over the maximal torus of the group $G$ with respect to the Haar measure, $[d\mu]_G$, associated with the group $G$.  The Hilbert series, as defined by Eq.~\eqref{eq:hilbertdef}, is a polynomial in the spurions such that the coefficient of different powers of the spurions counts the number of invariants under the group $G$.\footnote{The invariants are counted using the character orthogonality relation: 
\begin{align}
\oint [d\mu]_G ~\chi_{R}~\chi_{R^{\prime}}=\delta_{RR^{\prime}},
\end{align}
where $\chi_{R}$ and $\chi_{R^{\prime}}$ are characters of irreducible representations, $R$ and $R^{\prime}$, of $G$. When $R^{\prime}$ is a trivial singlet representation, using $\chi_{\text{singlet}}=1$, we have
\begin{equation}
\oint [d\mu]_G ~\chi_{R}= 1, \hspace{0.1in} \text{iff} ~\chi_R = \chi_{\text{singlet}}.
\end{equation}
Therefore, using the definition of the plethystic exponentials, the Hilbert series we use, which counts the invariants under group $G$ is 
\begin{equation}
HS = \oint [d\mu]_G~ PE.
\end{equation}
}
For further details, we refer to~\cite{Jenkins:2009dy, Hanany:2010vu, Feng:2007ur, Henning:2015daa, Henning:2015alf}.

Using the machinery of the Hilbert series, we can construct all possible operators $\mathcal{O}_k$ in Eq.~(\ref{lagrangian}).  The characters for $E$, $B$, $\psi$, $\psi^\dagger$, and $s$ are (note that $P_0$, $P_\perp$, $\mathcal{D}_t$, and $\mathcal{D}_\perp$ are defined in Eqs.~\eqref{P0def} and~\eqref{Pperpdef} and in the text thereafter):
\begin{eqnarray}
\label{chiE}
\chi_E &=& P_0 P_\perp\chi^C_E \left( \chi_{\bf 3}^{SO(3)} -  \mathcal{D}_\perp \chi_{\bf 3}^{SO(3)} + \mathcal{D}_\perp^2 \right) ,\\
\label{chiB}
\chi_B&=& P_0 P_\perp \chi^C_B  \left( \chi_{\bf 3}^{SO(3)} -  \mathcal{D}_\perp \right) , \\
\label{chipsi}
\chi_\psi &=& P_0 P_\perp \chi^C_\psi \chi_{\bf 2}^{SU(2)} (1-\mathcal{D}_t) , \\
\label{chipsid}
\chi_{\psi^\dagger} &=& P_0 P_\perp \chi^{C}_{\psi^\dagger}  \chi_{\bf 2}^{SU(2)} (1-\mathcal{D}_t) , \\
\chi_s &=& \chi_{\bf 3}^{SO(3)} \chi_{\bf 3}^{SU(2)} ,
\end{eqnarray}
where $\chi_{\bf 3}^{SO(3)}$, $\chi_{\bf 2}^{SU(2)}$, and $\chi_{\bf 3}^{SU(2)}$ are the characters for a ${\bf 3}$ of $SO(3)$, a ${\bf 2}$ of $SU(2)$, and a ${\bf 3}$ of $SU(2)$, respectively.  Explicit expressions for these characters can be found in Appendix~\ref{Haars}.  The characters $\chi^C$ represent the way $E$, $B$, $\psi$, and $\psi^\dagger$ are charged under the external gauge field.  For example, if the fermion has color, then $\chi^C_E$ and $\chi^C_B$ are both the characters for the adjoint representation of $SU(3)$, and $\chi^C_\psi$ ($\chi^C_{\psi^\dagger}$) is the character for the fundamental (antifundamental) representation of $SU(3)$.  
$P_0$ and $P_\perp$ generate all symmetric products of temporal and spatial derivatives, respectively:
\begin{eqnarray}
\label{P0def}
P_0 &\equiv& \exp\left[ \displaystyle \sum_{n=1}^\infty \frac{\mathcal{D}_t^n}{n}  \right] = \frac{1}{1-\mathcal{D}_t} , \\
\label{Pperpdef}
P_\perp &\equiv& \exp\left[ \displaystyle \sum_{n=1}^\infty \frac{\mathcal{D}_\perp^n}{n} \chi_{\bf 3}^{SO(3)}(z^n) \right] = \frac{1}{(1-z\mathcal{D}_\perp)(1-\mathcal{D}_\perp)(1-\mathcal{D}_\perp/z)} ,
\end{eqnarray}
where $\mathcal{D}_t$ and $\mathcal{D}_\perp$ are the spurions that correspond to time and spatial derivatives in the operator, respectively.  The characters in Eqs.~(\ref{chiE}) - (\ref{chipsid}) take the form they do so as to remove terms that are zero according to the equations of motion for the external gauge field, where we choose to construct operators with $\partial \bm{B}/\partial t$, in lieu of $\nabla \times \bm{E}$, according to Eq.~\eqref{gaugeEOM}. We note that, we have to add back $\mathcal{D}_{\perp}^{2}$ in Eq.~\eqref{chiE}, to enforce the constraint that $\nabla \cdot \left(\nabla \times \bm{E}\right)=0$.  Without it, Hilbert series will erroneously subtract off $\nabla \cdot \left(\nabla \times \bm{E}\right)$, which was not there to begin with before the subtraction.

The general Hilbert series for a theory with a heavy fermion is
\begin{eqnarray}
HS &=&  \oint [d\mu]_{SO(3)} \oint [d\mu]_{SU(2)}  \oint [d\mu]_{C}~ \frac{1}{P_0 P_\perp} ~PE_E ~PE_B~ PE_\psi ~PE_{\psi^\dagger} ~PE_s   .
\end{eqnarray}
The bosonic plethystic exponential, i.e., Eq.~\eqref{pebosons}, is used for $E$, $B$, and $s$, while the fermionic one, i.e., Eq.~\eqref{pefermions}, is used for $\psi$ and $\psi^\dagger$. The expressions for the Haar measures can be found in Appendix~\ref{Haars}.  The factor of $1/P_0P_\perp$ removes operators that are total time derivatives and total spatial derivatives.\footnote{One can justify introducing the factor of $1/P_0P_\perp$,  by noting that one can always choose a basis of operators where no time or spatial derivatives act on $\psi^{\dagger}$, using integration by parts. This procedure should remove the $P_{0}P_{\perp}$ in the definition of the character for $\psi^\dagger$.   The factor $1/P_0P_\perp$ can also be justified using differential forms, as discussed in Ref.~\cite{Henning:2015alf}.}  This method, however, will over-subtract operators that are total derivatives, but which have already been subtracted by the equations of motion.  Thus, this Hilbert series will, in general, produce some terms with negative signs, all of which are redundant operators, and can be ignored.
One can expand the plethystic exponentials for $\psi$ and $\psi^\dagger$ to first order, and perform the $SU(2)$ integral by hand, which results in the Hilbert series for the operators $\mathcal{O}_k$ in Eq.~(\ref{lagrangian}):
\begin{equation}
HS =  \oint [d\mu]_{SO(3)}   \oint [d\mu]_{C}~ \frac{P_\perp}{P_0 } (1+s\chi_{\bf 3}^{SO(3)}) ~\chi^{C}_{\psi^{\dagger}} \chi^{C}_\psi ~PE_E ~PE_B .
\end{equation}
Explicit expressions for the Hilbert series NRQED and NRQCD/HQET will be given in Sections~\ref{secNRQED} and~\ref{secNRQCD}, respectively, including discussions on how to impose invariance under parity and time reversal.

\section{NRQED}
\label{secNRQED}

In NRQED, the relevant gauge symmetry group is $U(1)$.  Here, $\chi^C_E =\chi^C_B = 1$, since photons do not have any $U(1)$ charge, and $\chi^C_\psi = \chi^{U(1)}_\psi = 1/\chi^{U(1)}_{\psi^\dagger}$.  Because of this, we have
\begin{equation}
 \oint [d\mu]_{U(1)}~ \chi^{U(1)}_{\psi^\dagger} \chi^{U(1)}_\psi =  \oint [d\mu]_{U(1)} =1
\end{equation} 
The Hilbert series for $\mathcal{O}_k$ in Eq.~(\ref{lagrangian}) in NRQED is
\begin{eqnarray}
HS &=&  \oint [d\mu]_{SO(3)} ~  \frac{P_\perp}{P_0 } (1+s\chi_{\bf 3}^{SO(3)})  ~PE_E ~PE_B  .
\end{eqnarray}
Again, we ignore any negative terms generated by this Hilbert series, since they are both total derivatives and related to other operators by the equations of motion, as discussed in Section~\ref{secHS}.  Since parity is a symmetry of QED, one can demand that $\mathcal{O}_k$ respects parity by requiring that it is composed of any number of parity-even objects, i.e., $\mathcal{D}_t$, $B$, and $s$, and an even number of parity-odd objects, i.e., $\mathcal{D}_\perp$ and $E$.  This can be automated without explicitly constructing the operators $\mathcal{O}_k$ by hand. 

The output for this Hilbert series for dimensions 5, 6, 7, and 8, before imposing invariance under time reversal, is
\begin{eqnarray}
HS_{d=5} &=& \mathcal{D}_\perp^2 + sB , \\
HS_{d=6} &=& 2E\mathcal{D}_\perp + sE\mathcal{D}_\perp ,  \\
HS_{d=7} &=& \mathcal{D}_\perp^4 + E^2 + B^2 + B \mathcal{D}_\perp^2 + 5 sB\mathcal{D}_\perp^2 , \\
HS_{d=8} &=& sB^2 \mathcal{D}_t + sE^2 \mathcal{D}_t + 2EB\mathcal{D}_\perp + 3sE\mathcal{D}_\perp^3 + 5E\mathcal{D}_\perp^3+ 7 sEB\mathcal{D}_\perp .
\end{eqnarray}
While the Hilbert series can count the number of operators that are invariant under the given symmetries, it does not say how the indices within each operator are contracted.  In general, this needs to be done by hand.  
To do this, we choose to organize operators according to what objects the derivatives are acting on. 
$E$ and $B$ have no electric charge, so derivatives acting on $E$ and $B$ are only partial derivatives.  
As such, objects of the form $[\partial_t ... \partial_\perp E]$ and $[\partial_t ... \partial_\perp B]$ are Hermitian, where the square brackets indicate that the derivatives only act on $E$ or $B$. 
So as not to introduce terms like $\nabla \cdot {\bf B}$ and $\nabla \times {\bf E}$, we require that the $SO(3)$ index of $B$ cannot be be symmetric with any index of $\partial_\perp$  acting on it, and the index on $E$ must be symmetric with the index of any $\partial_\perp$ acting on it.  
Because $\psi$ does have electromagnetic charge, the derivatives acting on it are covariant derivatives.  Only spatial derivatives can act on $\psi$, due to the equations of motion, and we use anticommutator brackets $\{A,B\}\equiv AB + BA$ to construct fully Hermitian operators $\mathcal{O}$. 
One can impose invariance under time reversal by hand, as shown in Table~\ref{NRQED} for $d$ = 5, 6, 7, and 8.   $T$-even operators are those with any number of $T$-even objects, i.e., $E$ and $\partial_\perp$, and an even numbers of $T$-odd objects, i.e., $\partial_t$, $iD_\perp$, $B$, and $s$. 

\begin{figure}[h!]
\centering
\includegraphics[width=.75\textwidth]{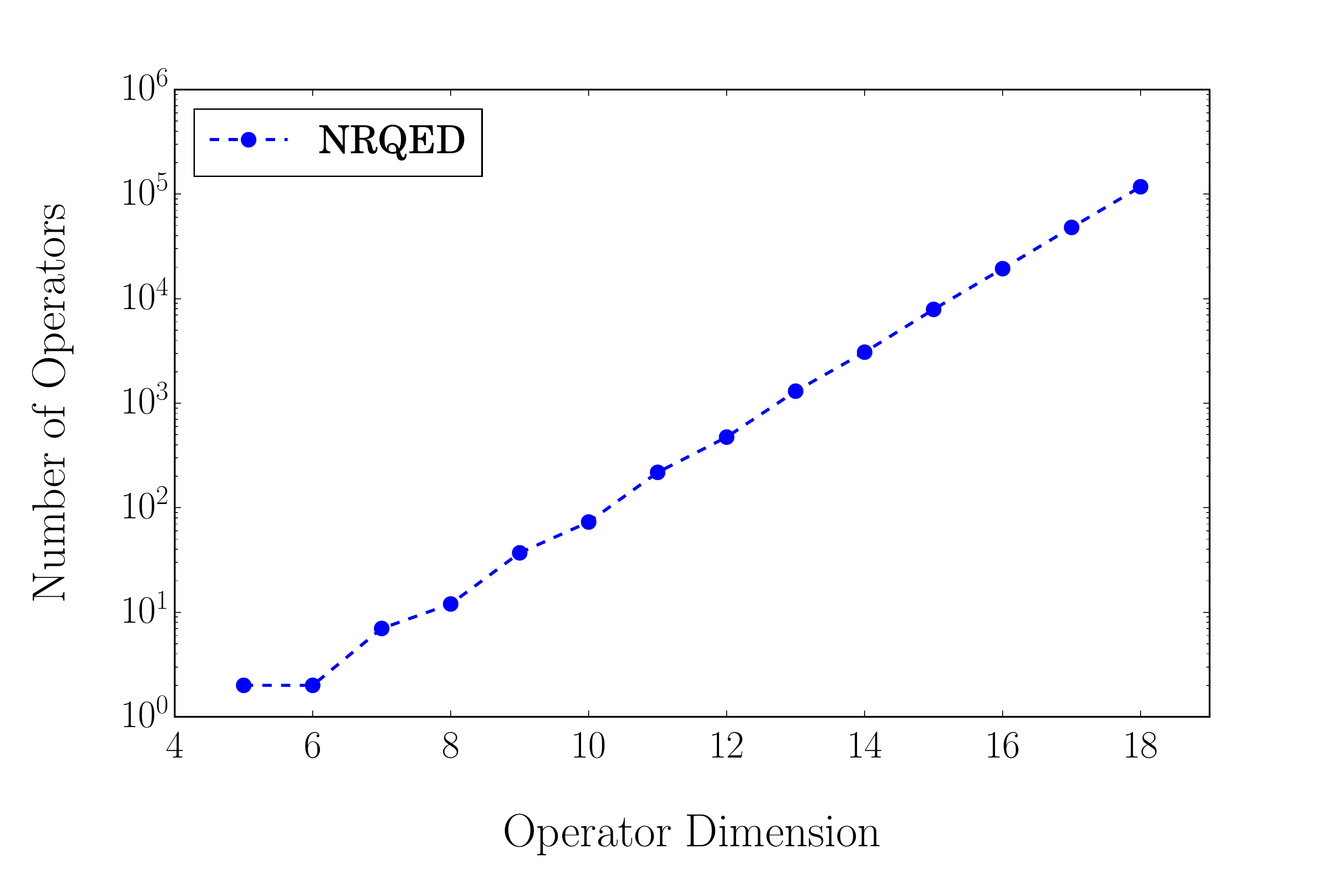}
\caption{The total number of rotationally-invariant operators in NRQED, which are even under parity and time reversal, as a function of the operator dimension $d$. Explicit form of the operators for $d=5,6,7,8$ can be found in Table~\ref{NRQED}. }
\label{QEDfig}
\end{figure}

In the special case of NRQED, where the group is abelian, there is a method to impose $T$ invariance that is easily automated.  This is done by modifying the Hilbert series to distinguish those spatial derivatives $\partial_\perp$ acting only on $E$ and $B$ from the spatial derivatives $iD_\perp$ that act on $\psi$. Here, the former ones are always $T$-even, while the latter are always $T$-odd.  This results in:
\begin{eqnarray}
HS_{d=5} &=& \mathcal{D}_\perp^2 + sB , \\
HS_{d=6} &=& E\mathcal{D}_\perp + sE\mathcal{D}_\perp , \\
HS_{d=7} &=& \mathcal{D}_\perp^4 + E^2 + B^2 + B \mathcal{D}_\perp^2 + 3 sB\mathcal{D}_\perp^2 , \\
HS_{d=8} &=& sB^2 \mathcal{D}_t + sE^2 \mathcal{D}_t + EB\mathcal{D}_\perp + 2sE\mathcal{D}_\perp^3 + 3E\mathcal{D}_\perp^3+ 4 sEB\mathcal{D}_\perp .
\end{eqnarray}
This method agrees with the result when explicitly constructing operators and selecting by hand only those that are $T$-even, and it agrees with the lists of operators up to and including $d=8$ in Refs.~\cite{Caswell:1985ui, Kinoshita:1995mt, Manohar:1997qy, Hill:2012rh}.  It is straight forward, using the Hilbert series as a guide, to explicitly list operators for $d>8$.  We show in Fig.~\ref{QEDfig} the total number of operators in NRQED up to $d=18$, when $T$ invariance is imposed, and list the total number of operators in Table~\ref{NRQEDnums}.

\begin{table}[h!tbp]
\centering
\begin{tabular}{| c || c | c | c | c | c | c | c | c | c | c | c | c | c | c | }
\hline
Mass dimension ($d$) & 5 & 6 & 7 & 8 & 9 & 10 & 11 & 12 & 13 & 14 & 15 & 16 & 17 & 18 \\ \hline  
Number of operators & 2 & 2 & 7 & 12 & 37 & 73 & 218 & 474 & 1303 & 3077 & 7896 & 19359 & 48023 & 117625 \\ \hline
\end{tabular}
\caption{The total number of effective operators in NRQED with mass dimension $d$ up to and including $d=18$, which are invariant under parity and time reversal transformations. }
\label{NRQEDnums}
\end{table}

\begin{table}[h!tbp]
\footnotesize
\centering
\resizebox{1 \textwidth}{!}{
\begin{tabular}{| c | c | c | c |}
\hline
Order & HS & $T$ even & $T$ odd \\ \hline \hline 

\multirow{2}{*}{$\frac{1}{m}$} & $\mathcal{D}_\perp^2$ & $(\iD)^2 $ & \\ \cline{2-4}
					     & $sB$ & $ s^i  B^j \delta_{ij} $ & \\ \hline \hline
					     
\multirow{2}{*}{$\frac{1}{m^2}$} & $2 E \mathcal{D}_\perp$ & $ [\p^i E^j] \delta_{ij}$ & $\{E^i,\iD^j\} \delta_{ij}$  \\  \cline{2-4}
					        							
						& $ sE \mathcal{D}_\perp$ & $s^i\{E^j,\iD^k\}\epsilon_{ijk}$  &  \\ \hline \hline

\multirow{7}{*}{$\frac{1}{m^3}$} & $\mathcal{D}_\perp^4$  & $(\iD)^4$  & \\ \cline{2-4} 
					        &  $E^2$ & $E^2$ &   \\ \cline{2-4}	
					        	& $B^2$ & $B^2$ &  \\ \cline{2-4}
						& $B \mathcal{D}_\perp^2$ & $\{ [\p^i B^j], \iD^k \}\epsilon_{ijk}$ & \\ \cline{2-4}
						& \multirow{3}{*}{$5sB \mathcal{D}_\perp^2$} & $\{ s^iB^j,(\iD^k)^2 \} \delta_{ij}$ & $\{ s^i [\p^j B^k], \iD^l \} \delta_{ij}\delta_{kl}$  \\ 	
												   & & $\{ s^iB^j,\iD^k \iD^l \} (\delta_{ik}\delta_{jl} +\delta_{il}\delta_{jk} )$  & $\{ s^i [\p^j B^k], \iD^l \} \delta_{ik}\delta_{jl}$  \\ 
												   & &  $s^i [\p^2 B^j] \delta_{ij}$ &  \\ \hline \hline
												   
\multirow{16}{*}{$\frac{1}{m^4}$} & $sB^2 \mathcal{D}_t$ & $s^i B^j [\partial_t B^k] \epsilon_{ijk}$ &  \\  \cline{2-4}
						  & $sE^2 \mathcal{D}_t$ & $s^i E^j [\partial_t E^k] \epsilon_{ijk}$ &  \\  \cline{2-4}
						  & $2 EB\mathcal{D}_\perp$ & $\{ E^i B^j, \iD^k \}\epsilon_{ijk}$ & $E^i [\p^j B^k]\epsilon_{ijk}$ \\  \cline{2-4} 
						  & \multirow{2}{*}{$3sE \mathcal{D}_\perp^3$} & $\{ s^i E^j, \iD^k (\iD)^2 \}\epsilon_{ijk}$ & $\{ s^i [\p^j E^k], \iD^l \iD^m \}(\epsilon_{ijl}\delta_{km} + \epsilon_{ijm}\delta_{kl}+\epsilon_{ikl}\delta_{jm}+\epsilon_{ikm}\delta_{jl})$ \\
												     & & $\{ s^i [\p^j \p^k E^l], \iD^m\}(\epsilon_{ijm}\delta_{kl}+\epsilon_{ikm}\delta_{jl}+\epsilon_{ilm}\delta_{jk})$ & \\ \cline{2-4} 
						  & \multirow{3}{*}{$5E \mathcal{D}_\perp^3$} & $\{ [\p^i E^j], (\iD)^2 \}\delta_{ij}$ & $\{ E^i, \iD^j \iD^k \iD^l \}(\delta_{ij}\delta_{kl}+\delta_{ik}\delta_{jl}+\delta_{il}\delta_{jk})$ \\
						  						    & &  $\{ [\p^i E^j], \iD^k \iD^l \}(\delta_{ik}\delta_{jl}+\delta_{il}\delta_{jk})$ & $\{ [\p^i \p^j E^k], \iD^l\}(\delta_{ij}\delta_{kl}+\delta_{ik}\delta_{jl}+\delta_{il}\delta_{jk})$ \\
						  						    & &  $[\p^i \p^j \p^k E^l](\delta_{ij}\delta_{kl}+\delta_{ik}\delta_{jl}+\delta_{il}\delta_{jk})$ &  \\  \cline{2-4} 
						  & \multirow{4}{*}{$7sEB \mathcal{D}_\perp$} & $s^i E^j [\p^k B^l]\delta_{ik}\delta_{jl}$ & $\{ s^i E^j B^k, \iD^l\} \delta_{ij} \delta_{kl} $ \\
						  						    & & $s^i E^j [\p^k B^l]\delta_{il}\delta_{jk}$ & $\{ s^i E^j B^k, \iD^l\} \delta_{ik} \delta_{jl} $ \\	
												    & & $s^i B^j [\p^k E^l] (\delta_{ik}\delta_{jl}+\delta_{il}\delta_{jk})$ & $\{ s^i E^j B^k, \iD^l\} \delta_{il} \delta_{jk} $ \\	
												    & & $s^i B^j [\p^k E^l] \delta_{ij}\delta_{kl}$ & \\	 \hline					    
						  
\end{tabular}
}
\caption{The output of the NRQED Hilbert series for mass dimensions $d=5, 6, 7,$ and 8.  We list the possible Hermitian combinations of these operators, distinguishing between those that are even and odd under time reversal, where $i,j,k,l,m,...$ signify $SO(3)$ indices.  Note that these Hermitian operators $\mathcal{O}$ are those in the bilinear $\psi^\dagger \mathcal{O}\psi$, and the square brackets indicate that the derivative acts only on the object in the square bracket.  Also, in the special case of NRQED, time-reversal symmetry can be imposed in an automated way, without constructing Hermitian operators by hand; see the text at the end of Section~\ref{secNRQED} for details.   }
\label{NRQED}
\end{table}

\section{NRQCD/HQET}
\label{secNRQCD}

The construction of the Hilbert series for NRQCD/HQET is very similar to that of NRQED, where now $\chi^C_E =\chi^C_B = \chi^{SU(3)}_{\bf 8}$,  $\chi^C_\psi = \chi^{SU(3)}_{\bf 3}$, and $\chi^C_{\psi^\dagger} = \chi^{SU(3)}_{\bf \bar{3}}$.  
The Hilbert series for the operators $\mathcal{O}_k$ in Eq.~(\ref{lagrangian}) is 
\begin{eqnarray}
HS &=&  \oint [d\mu]_{SO(3)} \oint [d\mu]_{SU(3)} ~  \frac{P_\perp}{P_0 } (1+s\chi_{\bf 3}^{SO(3)}) ~\chi_{\bf 3}^{SU(3)} ~\chi_{\bf \bar{3}}^{SU(3)} ~PE_E ~PE_B . 
\end{eqnarray}
When invariance under parity is imposed, the output from this Hilbert series for operators of mass dimension $d$ = 5, 6, 7, and 8 is:
\begin{eqnarray}
HS_{d=5} &=& \mathcal{D}_\perp^2 + sB , \\
HS_{d=6} &=& 2E\mathcal{D}_\perp + sE\mathcal{D}_\perp , \\
HS_{d=7} &=& \mathcal{D}_\perp^4 + 2E^2 + 2B^2 + sE^2 + sB^2 + B\mathcal{D}_\perp^2 + 5sB\mathcal{D}_\perp^2 , \\
HS_{d=8} &=& B^2 \mathcal{D}_t + E^2 \mathcal{D}_t + 2sB^2 \mathcal{D}_t + 2sE^2 \mathcal{D}_t + 6EB\mathcal{D}_\perp + 3sE\mathcal{D}^3_\perp + 5E\mathcal{D}^3_\perp + 21 s EB\mathcal{D}_\perp .
\end{eqnarray}
Unlike NRQED, we have not found an automated way to implement invariance under time reversal in NRQCD/HQET,
 because $T$ acts as an anti-unitary operator, and counting $T$-invariant operators requires keeping track of factors $i$ while constructing Hermitian operators.  This is not an issue when constructing invariants in NRQED, since it has an abelian $U(1)$ symmetry, but when the group is non-abelian, like $SU(3)$, the algebra's structure constants, e.g., $f_{abc}$, bring with them a factor of $i$, and imposing $T$-symmetry is no longer straight-forward.    

We take the output from this Hilbert series and explicitly contract indices by hand, separating those that are even and odd under time reversal.  
The prescription is very close to the one we used in NRQED.  We choose to suppress color indices, and express  $E = E_a T^a$ and $B=B_aT^a$, where $T^a$ are the eight generators of $SU(3)$, which satisfy: 
\begin{eqnarray}
\label{anti}
[T^a, T^b] &=& i f^{abc} T_c , \\
\label{sym}
\{T^a, T^b\} &=& \frac{1}{3} \delta^{ab} + d^{abc}T_c .
\end{eqnarray}
We utilize the following notation, where  the letters $i, j, k, l, m, ...$ are used for $SO(3)$ indices, and the letters $a, b, c, ...$ are used to signify the $SU(3)$ generators:
\begin{equation}
\psi^\dagger[D_\perp^i E^j]_a \delta_{ij} T^a \psi \equiv \psi^\dagger \left( [\partial_\perp^i E^j_a] + g(A_\perp)^{ib} E^{jc} f_{abc} \right) \delta_{ij} T^a \psi ,
\end{equation} 
where $A^{\mu}\equiv A^{\mu}_{a}T^{a}$ is the gauge field.  
When there are two $SU(3)$ generators in an operator, one can use the relation that results in adding Eqs.~(\ref{anti}) and (\ref{sym}) together:
\begin{equation}
\label{TaTb}
T^a T^b = \frac{1}{6} \delta^{ab} + \frac{1}{2}\left( d^{abc}T_c +  i f^{abc} T_c \right) .
\end{equation}
From this, one can see that, for example, the operator $E^2$ in the Hilbert series can be contracted in two ways:
\begin{equation}
\label{E2}
E^i_a E^j_b \delta_{ij} T^a T^b~~ \rightarrow~~ E^i_a E^j_b \delta_{ij}\delta^{ab} \hspace{0.1in} \text{and} \hspace{0.1in} E^i_a E^j_b \delta_{ij} d^{abc} T_c .
\end{equation}
A third contraction with $f_{abc}$ is completely antisymmetric in $a,b,c$, which results in an operator equal to zero, in this case.  Finally, it should be noted that $f^{abc}$ should be thought of a odd under time reversal.\footnote{This can be heuristically understood by noting that color is an internally symmetry, and must therefore be invariant under spacetime transformations.  Therefore, the matrix multiplication between two $SU(3)$ generators must be even under time reversal, which requires  $T^{-1}if^{abc}T = if^{abc}$, and $f^{abc}$ can be therefore be thought of as $T$-odd. }  The two contractions in Eq.~\eqref{E2}  give rise to different matrix elements, since the contraction of color indices would be different.  A complete list of NRQCD/HQET operators can be found in Table~\ref{NRQCD} for $d\leq 8$.  Extending the list to higher orders would be straight-forward. 

Our results agree with those in Ref.~\cite{Manohar:1997qy} for NRQCD/HQET operators up to and including operators of order $1/m^3$.  However, we find a different number compared to Refs.~\cite{Mannel:2010wj, Gunawardana:2017zix} for operators at order $1/m^3$ and $1/m^4$.  Specifically, Refs.~\cite{Mannel:2010wj, Gunawardana:2017zix} claim there are 9 operators at $1/m^3$, and 18 operators at $1/m^4$, while we find 11 and 25, respectively.  These discrepancies are consistent with the possibility that Refs.~\cite{Mannel:2010wj, Gunawardana:2017zix} count only once the two symmetric terms, i.e., contractions with $\delta^{ab}$ and $d^{abc}$, in Eq.~\eqref{TaTb}.

\begin{table}[h!]
\footnotesize
\centering
\resizebox{1 \textwidth}{!}{
\begin{tabular}{| c | c | c | c |}
\hline
Order & HS & $T$ even & $T$ odd \\ \hline \hline 

\multirow{2}{*}{$\frac{1}{m}$} & $\mathcal{D}_\perp^2$ & $(\iD)^2 $ & \\ \cline{2-4}
					     & $sB$ & $ s^i  B^j_a \delta_{ij} T^a $ & \\ \hline \hline
					     
\multirow{2}{*}{$\frac{1}{m^2}$} & $2 E \mathcal{D}_\perp$ & $ [\D^i E^j]_a \delta_{ij}T^a$ & $\{E^i_a,\iD^j\} \delta_{ij}T^a$  \\  \cline{2-4}

						& $ sE \mathcal{D}_\perp$ & $s^i\{E^j_a,\iD^k\}\epsilon_{ijk}T^a$  &  \\ \hline \hline
						
\multirow{12}{*}{$\frac{1}{m^3}$} & $\mathcal{D}_\perp^4$  & $(\iD)^4$  & \\ \cline{2-4} 
					        &  \multirow{2}{*}{$2E^2$} & $E^i_a E^j_b \delta_{ij} d^{abc} T_c $ &   \\ 
					        					     & & $E^i_a E^j_b \delta_{ij} \delta^{ab}$ & \\ \cline{2-4}	
					        	&  \multirow{2}{*}{$2B^2$} & $B^i_a B^j_b \delta_{ij} d^{abc} T_c $ &   \\ 
					        					     & & $B^i_a B^j_b \delta_{ij} \delta^{ab}$ & \\ \cline{2-4}	
						& $sE^2$ & $s^i E^j_a E^k_b \epsilon_{ijk} f^{abc}T_c$ &  \\	\cline{2-4}	
						& $sB^2$ & $s^i B^j_a B^k_b \epsilon_{ijk} f^{abc}T_c$ &  \\	\cline{2-4}	
						& $B\mathcal{D}_\perp^2$ & $\{ [\D^i B^j]_a, \iD^k \}\epsilon_{ijk}T^a$ &  \\	\cline{2-4}	
						& \multirow{3}{*}{$5sB \mathcal{D}_\perp^2$} & $\{ s^iB^j_a,(\iD^k)^2 \} \delta_{ij}T^a$ & $\{ s^i [\D^j B^k]_a, \iD^l \} \delta_{ij}\delta_{kl}T^a$  \\ 	
												   & & $\{ s^iB^j_a,\iD^k \iD^l \} (\delta_{ik}\delta_{jl} +\delta_{il}\delta_{jk} )T^a$  & $\{ s^i [\D^j B^k]_a, \iD^l \} \delta_{ik}\delta_{jl}T^a$  \\ 
												   & &  $s^i [\D^2 B^j]_a \delta_{ij}T^a$ &  \\ \hline \hline		
\multirow{30}{*}{$\frac{1}{m^4}$}	& $B^2\mathcal{D}_t$ & $B_a^i[D_t B^j]_b\delta_{ij} f^{abc}T_c$ & \\  \cline{2-4}
						& $E^2\mathcal{D}_t$ & $E_a^i[D_t E^j]_b\delta_{ij}f^{abc}T_c$ &    \\  \cline{2-4}
						& \multirow{2}{*}{$2sB^2 \mathcal{D}_t$}  &  $s^i  B^j_a [D_t B^k]_b \epsilon_{ijk}\delta^{ab}$ &   \\  
										     	     & &  $s^i  B^j_a [D_t B^k]_b \epsilon_{ijk} d^{abc}T_c$  & \\ \cline{2-4}
						& \multirow{2}{*}{$2sE^2 \mathcal{D}_t$}  &  $s^i  E^j_a [D_t E^k]_b \epsilon_{ijk}\delta^{ab}$ & \\  
										     	     & &  $s^i  E^j_a [D_t E^k]_b \epsilon_{ijk}d^{abc}T_c$ & \\ \cline{2-4}	
						& \multirow{3}{*}{$6EB\mathcal{D}_\perp$}  & $\{ E^i_a B^j_b, \iD^k\}\epsilon_{ijk}\delta^{ab}$ & $\{ E^i_a B^j_b, \iD^k\}\epsilon_{ijk}f^{abc}T_c$ \\   
												& & $\{ E^i_a B^j_b, \iD^k\}\epsilon_{ijk}d^{abc}T_c$ & $E^i_a [\D^j B_b^k]_b\epsilon_{ijk}\delta^{ab}$ \\
										     	        & & $E^i_a [\D^j B^k]_b\epsilon_{ijk}f^{abc}T_c$ & $E^i_a [\D^j B^k]_b\epsilon_{ijk}d^{abc}T_c$ \\ \cline{2-4}
						& \multirow{2}{*}{$3sE \mathcal{D}_\perp^3$} & $\{ s^i E^j_a, \iD^k (\iD)^2 \}\epsilon_{ijk}T^a$ & $\{ s^i [\D^j E^k]_a, \iD^l \iD^m \}T^a(\epsilon_{ijl}\delta_{km} + \epsilon_{ijm}\delta_{kl}+\epsilon_{ikl}\delta_{jm}+\epsilon_{ikm}\delta_{jl})$ \\ 
												     & & $\{ s^i [\D^j \D^k E^l]_a, \iD^m\}T^a(\epsilon_{ijm}\delta_{kl}+\epsilon_{ikm}\delta_{jl}+\epsilon_{ilm}\delta_{jk})$ &  \\ \cline{2-4} 					        											   
						& \multirow{3}{*}{$5E \mathcal{D}_\perp^3$} & $\{ [\D^i E^j]_a, (\iD)^2 \}\delta_{ij}T^a$ & $\{ E^i_a, \iD^j \iD^k \iD^l \}T^a(\delta_{ij}\delta_{kl}+\delta_{ik}\delta_{jl}+\delta_{il}\delta_{jk})$ \\
						  						    & &  $\{ [\D^i E^j]_a, \iD^k \iD^l \}T^a(\delta_{ik}\delta_{jl}+\delta_{il}\delta_{jk})$ & $\{ [\D^i \D^j E^k]_a, \iD^l\}T^a(\delta_{ij}\delta_{kl}+\delta_{ik}\delta_{jl}+\delta_{il}\delta_{jk})$ \\
						  						    & &  $[\D^i \D^j \D^k E^l]_aT^a(\delta_{ij}\delta_{kl}+\delta_{ik}\delta_{jl}+\delta_{il}\delta_{jk})$ &  \\  \cline{2-4} 	
						& \multirow{11}{*}{$21sEB \mathcal{D}_\perp$} & $\{ s^i E^j_a B^k_b, \iD^l\} \delta_{ij} \delta_{kl}f^{abc}T_c $ & $\{ s^i E^j_a B^k_b, \iD^l\} \delta_{ij} \delta_{kl}d^{abc}T_c $ \\	
												     & & $\{ s^i E^j_a B^k_b, \iD^l\} \delta_{ik} \delta_{jl}f^{abc}T_c $ & $\{ s^i E^j_a B^k_b, \iD^l\} \delta_{ik} \delta_{jl} d^{abc}T_c$ \\
												     & & $\{ s^i E^j_a B^k_b, \iD^l\} \delta_{il} \delta_{jk}f^{abc}T_c $ & $\{ s^i E^j_a B^k_b, \iD^l\} \delta_{il} \delta_{jk} d^{abc}T_c$ \\
												     & & $s^i E^j_a [\D^k B^l]_b\delta_{ik}\delta_{jl}d^{abc}T_c$ & $\{ s^i E^j_a B^k_b, \iD^l\} \delta_{ij} \delta_{kl} \delta^{ab}$ \\
												     & & $s^i E^j_a [\D^k B^l]_b\delta_{il}\delta_{jk}d^{abc}T_c$ & $\{ s^i E^j_a B^k_b, \iD^l\} \delta_{ik} \delta_{jl} \delta^{ab}$ \\
												     & & $s^i E^j_a [\D^k B^l]_b\delta_{ik}\delta_{jl}\delta^{ab}$& $\{ s^i E^j_a B^k_b, \iD^l\} \delta_{il} \delta_{jk}\delta^{ab} $ \\
												     & & $s^i E^j_a [\D^k B^l]_b\delta_{il}\delta_{jk}\delta^{ab}$ & $s^i E^j_a [\D^k B^l]_b\delta_{ik}\delta_{jl}f^{abc}T_c$ \\
												     & & $s^i B^j_a [\D^k E^l]_b d^{abc}T_c(\delta_{ik}\delta_{jl}+\delta_{il}\delta_{jk})$ & $s^i E^j_a [\D^k B^l]_b\delta_{il}\delta_{jk}f^{abc}T_c$ \\
												     & & $s^i B^j_a [\D^k E^l]_b \delta_{ij}\delta_{kl}d^{abc}T_c$ & $s^i B^j_a [\D^k E^l]_b f^{abc}T_c (\delta_{ik}\delta_{jl}+\delta_{il}\delta_{jk})$ \\
												     & & $s^i B^j_a [\D^k E^l]_b \delta^{ab}(\delta_{ik}\delta_{jl}+\delta_{il}\delta_{jk})$ & $s^i B^j_a [\D^k E^l]_b \delta_{ij}\delta_{kl}f^{abc}T_c$ \\
												     & & $s^i B^j_a [\D^k E^l]_b \delta^{ab} \delta_{ij}\delta_{kl}$ & \\ \hline

\end{tabular}
}
\caption{Same as Table~\ref{NRQED}, but for NRQCD/HQET, separating those operators that are even and odd under time reversal.  See the text at the end of Section~\ref{secNRQCD} for a discussion regarding notation. }
\label{NRQCD}
\end{table}

\section{Discussion and Conclusions}
\label{concl}

We construct a Hilbert series for an effective theory with a single non-relativistic fermion in an external, and dynamical, gauge field, defining characters and using a method to subtract operators that are related to others via the equations of motion associated with the heavy fermion and the external gauge bosons, as well as integration by parts.  
We consider the specific examples of NRQED and NRQCD/HQET, where the heavy fermion has electric charge or color, respectively.  
Imposing invariance under parity can be easily automated.  Invariance under time reversal also can be automated in the case of NRQED, since it is an abelian theory, but not for NRQCD/HQET, in which case we separate $T$-even and -odd operators by hand.  For both effective theories, we construct explicit contractions for effective operators at dimensions $d\leq 8$ that are invariant under parity and time reversal transformations, as enumerated in Table~\ref{NRQED} for NRQED and Table~\ref{NRQCD} for NRQCD/HQET.  
In a theory with a nonrelativistic fermion, additional symmetry, e.g., reparameterization invariance~\cite{Luke:1992cs, Manohar:1997qy} or residual Lorentz symmetry~\cite{Heinonen:2012km}, is expected, in general. 
However, we do not impose such additional constraints, since it remains an open question regarding how to encode such requirements with Hilbert-series methods. 

Our results agree with those presented in Refs.~\cite{Caswell:1985ui, Kinoshita:1995mt, Manohar:1997qy, Hill:2012rh} for NRQED, which discuss operators up to and including $d=8$.  
The total number of operators in NRQED grows exponentially, as shown Fig.~\ref{QEDfig} and listed in Table~\ref{NRQEDnums} for mass dimension $d\leq 18$. 
When using a Hilbert series for NRQCD/HQET, we count a total of 2 operators each at orders $1/m$ and $1/m^2$, which agrees with Ref.~\cite{Mannel:1994kv}, 11 operators at $1/m^3$, which agrees with Ref.~\cite{Manohar:1997qy}, and 25 operators at $1/m^4$.  
However, at order $1/m^3$, other analyses claim that there are either 5~\cite{Dassinger:2006md}, or 9~\cite{Mannel:2010wj, Gunawardana:2017zix} total operators, and Refs.~\cite{Mannel:2010wj, Gunawardana:2017zix} claim there are a total of 18 operators at $1/m^4$.   
The differences between our results and those found in Refs.~\cite{Mannel:2010wj, Gunawardana:2017zix} can be explained by the existence of two symmetric $SU(3)$ color singlets for operators with two gauge bosons, as discussed at the end of Section~\ref{secNRQCD}.  It is possible that analyses using the results in Refs.~\cite{Dassinger:2006md, Mannel:2010wj, Gunawardana:2017zix}, may need to be reevaluated, e.g., Refs.~\cite{Gambino:2016jkc, Heinonen:2016cwm, Heinonen:2014dxa}. 

The authors of Refs.~\cite{Henning:2015daa, Henning:2015alf} discuss a connection between enumerating operators in a relativistic effective theory and the representations of the relativistic conformal group.  Here,  selecting only primary operators constructed out of tensor products of the conformal group's short representations correctly accounts for redundancies between operators via integration by parts and the equations of motion. 
We strongly suspect that our results can be reformulated in terms of the non-relativistic conformal group~\cite{Balasubramanian:2008dm, Goldberger:2014hca, Nishida:2007pj, Jensen:2014hqa, Pal:2017ntk}, and we take this up as future work.

{\it Note:} While this article was in review for publication, the authors of Ref.~\cite{Gunawardana:2017zix} updated their work, and their results now agree with our enumeration of NRQCD/HQET effective operators for $d\leq 8$.

\begin{acknowledgments}
We thank Aneesh Manohar and Jaewon Song for providing useful feedback.  We appreciate the authors of Ref.~\cite{Gunawardana:2017zix} pointing out some typographical errors in Table~\ref{NRQCD} and Adam Martin for pointing out a typo in our Appendix.  This work is supported in part by DOE grant \#DE-SC-0009919.
\end{acknowledgments}

\appendix

\section{Characters and Haar Measures}
\label{Haars}

The characters for irreducible representations needed in the NRQED and NRQCD/HQET Hilbert series for $U(1)$, $SO(3)$, $SU(2)$, and $SU(3)$ are
\begin{eqnarray}
\chi^{U(1)}(x) &=& x \\
\chi_{\bf 3}^{SO(3)}(z) &=& z^2 + 1 + \frac{1}{z^2} \\
\chi_{\bf 2}^{SU(2)}(y) &=& y + \frac{1}{y} \\
\chi_{\bf 3}^{SU(2)}(y) &=& y^2 + 1 + \frac{1}{y^2} \\
\chi_{\bf 3}^{SU(3)}(x_1, x_2) &=& x_2 + \frac{x_1}{x_2} + \frac{1}{x_1} \\
\chi_{\bf \bar{3}}^{SU(3)}(x_1, x_2) &=& x_1 + \frac{x_2}{x_1} + \frac{1}{x_2} \\
\chi_{\bf 8}^{SU(3)}(x_1, x_2) &=& x_1x_2 + \frac{x_2^2}{x_1} + \frac{x_1^2}{x_2} + 2 + \frac{x_1}{x_2^2} + \frac{x_2}{x_1^2} + \frac{1}{x_1x_2} 
\end{eqnarray} 
The contours integrals with respect to the Haar measures used in this analysis are
\begin{eqnarray}
\displaystyle \oint [d\mu]_{U(1)} &\equiv& \frac{1}{2\pi i} \oint_{|x|=1} \frac{1}{x}\\
\displaystyle \oint [d\mu]_{SO(3)} &\equiv& \frac{1}{2\pi i} \oint_{|z|=1} \frac{1}{2z} (1-z^2)\left(1-\frac{1}{z^2}\right) \\
\displaystyle \oint [d\mu]_{SU(2)} &\equiv& \frac{1}{2\pi i} \oint_{|y|=1}  \frac{1}{2y} (1-y^2)\left(1-\frac{1}{y^2}\right) \\
\displaystyle \oint [d\mu]_{SU(3)} &\equiv& \frac{1}{(2\pi i)^2} \oint_{|x_1|=1} \oint_{|x_2|=1} \frac{1}{6x_1x_2}(1-x_1x_2)\left(1-\frac{x_1^2}{x_2} \right)\left(1-\frac{x_2^2}{x_1} \right)\left(1-\frac{1}{x_1x_2} \right)\left(1-\frac{x_1}{x_2^2} \right)\left(1-\frac{x_2}{x_1^2} \right) \nonumber \\
\end{eqnarray}

\bibliography{bib}{}

\end{document}